\documentclass[a4paper]{jpconf}
\usepackage{amsmath}
\usepackage{amstext} 
\usepackage{cite}
\usepackage{graphicx}
\usepackage[breaklinks=true,colorlinks=true,linkcolor=blue,urlcolor=blue,citecolor=blue]{hyperref}
\usepackage{enumerate}
\usepackage{multirow}	
\usepackage{soul}       
\usepackage{ulem}       
\usepackage[table,xcdraw]{xcolor}     
\usepackage[hang]{footmisc}	
\usepackage{comment}		

\setstcolor{red}    

\begin{document}

\title{Polarising questions in the Force Concept Inventory}

\author{Anna Chrysostomou\textsuperscript{1}, 
Emanuela Carleschi\textsuperscript{1}, 
Alan S. Cornell\textsuperscript{1} and Wade Naylor\textsuperscript{1,2}} 
\address{\textsuperscript{1} Department of Physics, University of Johannesburg, PO Box 524, Auckland Park 2006, South Africa}
\address{\textsuperscript{2} Immanuel Lutheran College, PO Box 5025, Maroochydore 4558, Australia}
\ead{ecarleschi@uj.ac.za, annachrys97@gmail.com, acornell@uj.ac.za, naylorw@immanuel.qld.edu.au}

\begin{abstract}
The Force Concept Inventory (FCI) is a well-established physics education assessment tool used to evaluate students' comprehension of elementary mechanics principles. While it can be used to analyse the effectiveness of instruction if deployed as a pre- and post-test, we utilise the FCI here as pre-test only, to extract insights into first-year students' (mis)conceptions of Newtonian mechanics as they enter university. In this preliminary study, we tested 353 students enrolled at the University of Johannesburg in 2021, across six introductory physics courses. We focus on their responses to six ``polarising" questions, for which the presence of a correct and a mostly-correct answer allows for a clear demonstration of persistent misconceptions.
\end{abstract}




\section{Introduction \label{sec:1}}
\par Physics pedagogy has become a topic of interest in recent decades, with a particular focus on confronting persistent misconceptions generally held by incoming undergraduate students \cite{Hammer:1996,Obaidat:2008,Bayraktar:2009,Martin_Blas:2010,Savinainen:2008,Bani_Salameh:2016a,Bani_Salameh:2016b}. While the development of these misconceptions may arise from a number of sources beyond the control of the educator, Bani-Salameh has suggested that these flaws in understanding should be diagnosed as early as possible if effective teaching is to take place \cite{Bani_Salameh:2016a}. To identify these misconceptions, a number of diagnostic tools have been developed for the various physics sub-disciplines. A subset of these have been designed to gauge student comprehension at the onset of an introductory university course, and then to assess their progress at the end of the course. 

\par One such tool is the {\it Force Concept Inventory} (FCI) \cite{Hall:1985,Hest:1992, Hest:1998}, a 30-minute multiple-choice test comprised of 30 questions on elementary principles and applications of classical mechanics \textit{viz.} circular motion, Newton's three laws, etc.  The recommended means of administering this assessment tool is as a strictly closed-book test held at the beginning of the semester, to provide instructors with an indication of their students' baseline mechanics skills. Note that students are not expected to prepare for the assessment. This is known as the ``pre-test". Once the pre-test is completed it is not reviewed in class and students receive no feedback on their attempts. At the end of the semester the same FCI test is administered once again to students, without their prior knowledge of the re-testing. This is the ``post-test" phase. A common statistical measurement used to quantify student progress is the normalised gain $G$ \cite{Hake:1998}, defined as
\begin{equation}
G = \frac{\langle \% S_f \rangle  - \langle \% S_i \rangle }{100 - \langle \% S_i \rangle },
\label{eq:G}
\end{equation}
\noindent
where $\% S_f$ and $\% S_i$ are the final and initial FCI test scores, respectively.
\par Another way to interpret FCI data is to investigate individual questions, as recently performed by Alinea and Naylor \cite{Alinea:2017,Alinea:2015,Alinea:2020}. In these works, a breakdown of the type of response to each of the 30 questions demonstrated a ``polarisation" between a correct and a partially-correct answer within several questions (specifically, questions: 5, 11, 13, 18, 29, and 30). Students who choose the incorrect answer may unwittingly reveal that they suffer from one of the commonly held misconceptions outlined in references \cite{Martin_Blas:2010,Bani_Salameh:2016a, Bani_Salameh:2016b}. Our preliminary investigations conducted with the first-year engineering and physics students at the University of Johannesburg in 2020 \cite{Carleschi:2021} seemed to support this, where one particularly popular incorrect response was that ``a force in the direction of motion" was partially responsible for the action described in the question. The belief that ``motion requires an active force" was flagged as a misconception as identified in references \cite{Martin_Blas:2010,Bani_Salameh:2016a, Bani_Salameh:2016b}.

\par In our present study, we have performed this question-by-question analysis on the FCI pre-test responses from six first-year physics classes, with a total of $N=353$ students. We begin with a description of the student participants and the manner in which their responses were collected in section \ref{sec:2}. We then present our results in section \ref{sec:3}, with a focus on the polarising questions provided in table \ref{table2}. Conclusions and future avenues of pursuit are presented in\\ section \ref{sec:4}. 






\section{Methodology and data collection}
\label{sec:2}

\par Due to the ongoing nature of the COVID-19 pandemic, South African institutions of tertiary education were expected to function mostly $-$ if not entirely $-$ online. The delayed release of the 2020 Grade 12 results further derailed the 2021 academic programme, with first-year academic activities only beginning on March 8$^{\text{th}}$ at the University of Johannesburg, leading to a first semester shortened by several weeks.

\par To allow for first-year students to adjust to online learning, the authors of this study were forced to delay the deployment of the FCI test to the beginning of April. With the permissions of the various lecturers involved in each of the six courses that participated in this study (see table \ref{table1} for details), we launched the FCI test via the Blackboard interface and made it available to students for a total of five days, using the module page corresponding to each of the six courses. Through this platform, we were able to track student activity (to determine whether test-takers left the browser page during the course of the test), time their test attempt, and force submission after an allotted time. We were also able to make a test visible for a set period of time. For our data collection, we enabled these features to reduce cheating and to ensure students submitted their responses after 30 minutes. 


\begin{table}[t]
\caption{\label{table1} Course codes, their associated entry-level requirements (Grade 12 mathematics and physical sciences scores), FCI deployment period, number of student responses per class, and average scores for the six first-year classes involved in the 2021 FCI pre-testing. Note that Ext. refers to the extended courses, where students who do not meet entry-level requirements for introductory physics classes can complete a four-year Bachelor of Science in which the traditional first-year physics course is taught over a three-semester period.}
\begin{center}
\begin{tabular}{llllll}
\br
Course    &   Ent. \% & Ent. \%         &  Deployment               & Responses/ & Mean
\\
(Physics for...)  & Math. & Phys. Sci. &period & class & (\%) \\
\mr
PHYS1A1 (Majors)                     &  70    &  60     &13/04 - 17/04     & 19/70          & 33.2\\
PHY1EA1 (Physics Ext.: Sem1)       &  60    &   50    & 09/04 - 13/04     & 187/306        & 30.1         \\
PSFT0A1 (Education)                  &  50    &  50    & 05/04 - 09/04     & 62/91          & 29.9         \\
PHYG1A1 (Earth Sci.)             &  60    &   50    & 05/04 - 09/04     & 13/17          & 28.2         \\
PHYL1A1 (Life Sci.)              &  70    &  60     & 07/04 - 11/04     & 32/64          & 28.6         \\
PHE3LA1 (Life Sci. Ext.: Sem3) &  60    &   50    & 13/04 - 17/04     & 40/87          & 38.2         \\       
\br
\end{tabular}
\end{center}
\end{table}

\par Once the deployment period was over and each class had completed the assignment, the data was downloaded and processed within a single Excel  spreadsheet for each of the six individual classes. All data was anonymised in accordance with the requirements of the protection of personal information act.

\par In what follows we will present some preliminary analyses for FCI pre-test data for the 2021 cohort at the University of Johannesburg. The analyses break down the responses for each question to look for dominant misconceptions \cite{Martin_Blas:2010}. This can be very useful in gauging how an initial cohort and indeed different subgroups need different teaching pedagogies.

\section{Pre-test analyses and results
\label{sec:3}}


\begin{table}
\caption{\label{table2} A summary of the responses from the six participating classes to the polarising questions. Here, green [grey] and blue [light-grey] represent the correct and polarising choice \cite{Alinea:2020}, respectively, while orange [dark-grey] denotes responses more or equally popular.}
\begin{center}
\begin{tabular}{llllllll}
\br
\textbf{Course}        & \textbf{Answer}           & \textbf{Q5}                  & \textbf{Q11}                 & \textbf{Q13}                 & \textbf{Q18}                 & \textbf{Q29}                 & \textbf{Q30}                 \\
\mr
                 & \textbf{A}          & 10.5                         & 10.5                         & 15.8                         & 5.3                          & 0.0                          & 0.0                          \\
\textbf{}        & \textbf{B}          & \cellcolor[HTML]{81D41A}26.3 & 5.3                          & 5.3                          & \cellcolor[HTML]{81D41A}31.6 & \cellcolor[HTML]{81D41A}73.7 & 5.3                          \\
\textbf{}        & \textbf{C}          & \cellcolor[HTML]{FF8243}26.3 & \cellcolor[HTML]{ABCDEF}63.2 & \cellcolor[HTML]{ABCDEF}36.8 & 15.8                         & 5.3                          & \cellcolor[HTML]{81D41A}21.1 \\
\textbf{PHYS1A1} & \textbf{D}          & \cellcolor[HTML]{ABCDEF}15.8 & \cellcolor[HTML]{81D41A}15.8 & \cellcolor[HTML]{81D41A}26.3 & \cellcolor[HTML]{ABCDEF}21.1 & \cellcolor[HTML]{ABCDEF}10.5 & 5.3                          \\
\textbf{}        & \textbf{E}          & \cellcolor[HTML]{FF8243}21.1 & 0.0                          & 5.3                          & \cellcolor[HTML]{FF8243}21.1 & 0.0                          & \cellcolor[HTML]{ABCDEF}47.4 \\
\textbf{}        & \textbf{none}       & 0.0                          & 5.3                          & 10.5                         & 5.3                          & \cellcolor[HTML]{FF8243}10.5 & \cellcolor[HTML]{FF8243}21.1 \\
\textbf{}        & \textbf{total}      & 100.0                        & 100.0                        & 100.0                        & 100.0                        & 100.0                        & 100.0                        \\
\mr
                 & \textbf{A}          & \cellcolor[HTML]{FF8243}16.0 & \cellcolor[HTML]{FF8243}15.5 & 20.3                         & 5.3                          & \cellcolor[HTML]{FF8243}12.3 & 4.8                          \\
\textbf{}        & \textbf{B}          & \cellcolor[HTML]{81D41A}13.9 & \cellcolor[HTML]{FF8243}26.7 & \cellcolor[HTML]{FF8243}26.7 & \cellcolor[HTML]{81D41A}23.5 & \cellcolor[HTML]{81D41A}56.7 & \cellcolor[HTML]{FF8243}19.8 \\
\textbf{}        & \textbf{C}          & \cellcolor[HTML]{FF8243}40.6 & \cellcolor[HTML]{ABCDEF}36.9 & \cellcolor[HTML]{ABCDEF}23.5 & 16.0                         & 2.7                          & \cellcolor[HTML]{81D41A}13.4 \\
\textbf{PHY1EA1} & \textbf{D}          & \cellcolor[HTML]{ABCDEF}8.0  & \cellcolor[HTML]{81D41A}10.7 & \cellcolor[HTML]{81D41A}20.9 & \cellcolor[HTML]{ABCDEF}25.1 & \cellcolor[HTML]{ABCDEF}10.2 & 6.4                          \\
\textbf{}        & \textbf{E}          & \cellcolor[HTML]{FF8243}17.6 & 4.8                          & 3.2                          & 21.4                         & 3.2                          & \cellcolor[HTML]{ABCDEF}35.3 \\
\textbf{}        & \textbf{none}       & 3.7                          & 5.3                          & 5.3                          & 8.6                          & \cellcolor[HTML]{FF8243}15.0 & \cellcolor[HTML]{FF8243}20.3 \\
\textbf{}        & \textbf{total}      & 100.0                        & 100.0                        & 100.0                        & 100.0                        & 100.0                        & 100.0                        \\
\mr
                 & \textbf{A}          & \cellcolor[HTML]{FF8243}12.9 & \cellcolor[HTML]{FF8243}11.3 & \cellcolor[HTML]{FF8243}22.6 & 4.8                          & \cellcolor[HTML]{FF8243}16.1 & 4.8                          \\
\textbf{}        & \textbf{B}          & \cellcolor[HTML]{81D41A}14.5 & \cellcolor[HTML]{FF8243}40.3 & \cellcolor[HTML]{FF8243}27.4 & \cellcolor[HTML]{81D41A}21.0 & \cellcolor[HTML]{81D41A}50.0 & 11.3                         \\
\textbf{}        & \textbf{C}          & \cellcolor[HTML]{FF8243}41.9 & \cellcolor[HTML]{ABCDEF}25.8 & \cellcolor[HTML]{ABCDEF}19.4 & 16.1                         & 8.1                          & \cellcolor[HTML]{81D41A}14.5 \\
\textbf{PSFT0A1} & \textbf{D}          & \cellcolor[HTML]{ABCDEF}11.3 & \cellcolor[HTML]{81D41A}9.7  & \cellcolor[HTML]{81D41A}24.2 & \cellcolor[HTML]{ABCDEF}29.0 & \cellcolor[HTML]{ABCDEF}9.7  & 12.9                         \\
\textbf{}        & \textbf{E}          & \cellcolor[HTML]{FF8243}17.7 & \cellcolor[HTML]{FF8243}11.3 & 3.2                          & \cellcolor[HTML]{FF8243}27.4 & 4.8                          & \cellcolor[HTML]{ABCDEF}41.9 \\
\textbf{}        & \textbf{none}       & 1.6                          & 1.6                          & 3.2                          & 1.6                          & \cellcolor[HTML]{FF8243}11.3 & \cellcolor[HTML]{FF8243}14.5 \\
\textbf{}        & \textbf{total}      & 100.0                        & 100.0                        & 100.0                        & 100.0                        & 100.0                        & 100.0                        \\
\mr
                 & \textbf{A}          & 7.7                          & \cellcolor[HTML]{FF8243}7.7  & \cellcolor[HTML]{FF8243}23.1 & \cellcolor[HTML]{FF8243}15.4 & 7.7                          & 0.0                          \\
\textbf{}        & \textbf{B}          & \cellcolor[HTML]{81D41A}15.4 & \cellcolor[HTML]{FF8243}23.1 & \cellcolor[HTML]{FF8243}23.1 & \cellcolor[HTML]{81D41A}15.4 & \cellcolor[HTML]{81D41A}53.8 & 0.0                          \\
\textbf{}        & \textbf{C}          & \cellcolor[HTML]{FF8243}30.8 & \cellcolor[HTML]{ABCDEF}30.8 & \cellcolor[HTML]{ABCDEF}23.1 & \cellcolor[HTML]{FF8243}23.1 & 0.0                          & \cellcolor[HTML]{81D41A}7.7  \\
\textbf{PHYG1A1} & \textbf{D}          & \cellcolor[HTML]{ABCDEF}15.4 & \cellcolor[HTML]{81D41A}7.7  & \cellcolor[HTML]{81D41A}23.1 & \cellcolor[HTML]{ABCDEF}30.8 & \cellcolor[HTML]{ABCDEF}15.4 & \cellcolor[HTML]{FF8243}23.1 \\
\textbf{}        & \textbf{E}          & \cellcolor[HTML]{FF8243}23.1 & \cellcolor[HTML]{FF8243}7.7  & 0.0                          & 7.7                          & 0.0                          & \cellcolor[HTML]{ABCDEF}46.2 \\
\textbf{}        & \textbf{none}       & 7.7                          & \cellcolor[HTML]{FF8243}23.1 & 7.7                          & 7.7                          & \cellcolor[HTML]{FF8243}23.1 & \cellcolor[HTML]{FF8243}23.1 \\
\textbf{}        & \textbf{total}      & 100.0                        & 100.0                        & 100.0                        & 100.0                        & 100.0                        & 100.0                        \\
\mr
                 & \textbf{A}          & \cellcolor[HTML]{FF8243}25.0 & 9.4                          & 15.6                         & 3.1                          & \cellcolor[HTML]{FF8243}9.4  & 15.6                         \\
\textbf{}        & \textbf{B}          & \cellcolor[HTML]{81D41A}6.3  & 12.5                         & 18.8                         & \cellcolor[HTML]{81D41A}28.1 & \cellcolor[HTML]{81D41A}65.6 & 0.0                          \\
\textbf{}        & \textbf{C}          & \cellcolor[HTML]{FF8243}31.3 & \cellcolor[HTML]{ABCDEF}43.8 & \cellcolor[HTML]{ABCDEF}34.4 & 12.5                         & 0.0                          & \cellcolor[HTML]{81D41A}21.9 \\
\textbf{PHYL1A1} & \textbf{D}          & \cellcolor[HTML]{ABCDEF}12.5 & \cellcolor[HTML]{81D41A}15.6 & \cellcolor[HTML]{81D41A}25.0 & \cellcolor[HTML]{ABCDEF}31.3 & \cellcolor[HTML]{ABCDEF}6.3  & 6.3                          \\
\textbf{}        & \textbf{E}          & \cellcolor[HTML]{FF8243}21.9 & 6.3                          & 0.0                          & 18.8                         & 0.0                          & \cellcolor[HTML]{ABCDEF}34.4 \\
\textbf{}        & \textbf{none}       & 3.1                          & 12.5                         & 6.3                          & 6.3                          & \cellcolor[HTML]{FF8243}18.8 & \cellcolor[HTML]{FF8243}21.9 \\
\textbf{}        & \textbf{total}      & 100.0                        & 100.0                        & 100.0                        & 100.0                        & 100.0                        & 100.0                        \\
\mr
                 & \textbf{A}          & 7.5                          & \cellcolor[HTML]{FF8243}22.5 & 5.0                          & 5.0                          & \cellcolor[HTML]{FF8243}7.5  & 0.0                          \\
\textbf{}        & \textbf{B}          & \cellcolor[HTML]{81D41A}30.0 & 17.5                         & 17.5                         & \cellcolor[HTML]{81D41A}42.5 & \cellcolor[HTML]{81D41A}60.0 & 22.5                         \\
\textbf{}        & \textbf{C}          & \cellcolor[HTML]{FF8243}30.0 & \cellcolor[HTML]{ABCDEF}35.0 & \cellcolor[HTML]{ABCDEF}37.5 & 7.5                          & \cellcolor[HTML]{FF8243}10.0 & \cellcolor[HTML]{81D41A}32.5 \\
\textbf{PHE3LA1} & \textbf{D}          & \cellcolor[HTML]{ABCDEF}20.0 & \cellcolor[HTML]{81D41A}22.5 & \cellcolor[HTML]{81D41A}37.5 & \cellcolor[HTML]{ABCDEF}27.5 & \cellcolor[HTML]{ABCDEF}7.5  & 2.5                          \\
\textbf{}        & \textbf{E}          & 10.0                         & 0.0                          & 2.5                          & 15.0                         & 5.0                          & \cellcolor[HTML]{ABCDEF}30.0 \\
\textbf{}        & \textbf{none} & 2.5                          & 2.5                          & 0.0                          & 2.5                          & \cellcolor[HTML]{FF8243}10.0 & 12.5                         \\
\textbf{}        & \textbf{total}      & 100.0                        & 100.0                        & 100.0                        & 100.0                        & 100.0                        & 100.0            \\           
\br
\end{tabular}
\end{center}
\end{table}

\par
\begin{figure}
    \centering
    \scalebox{0.55}{
    \includegraphics{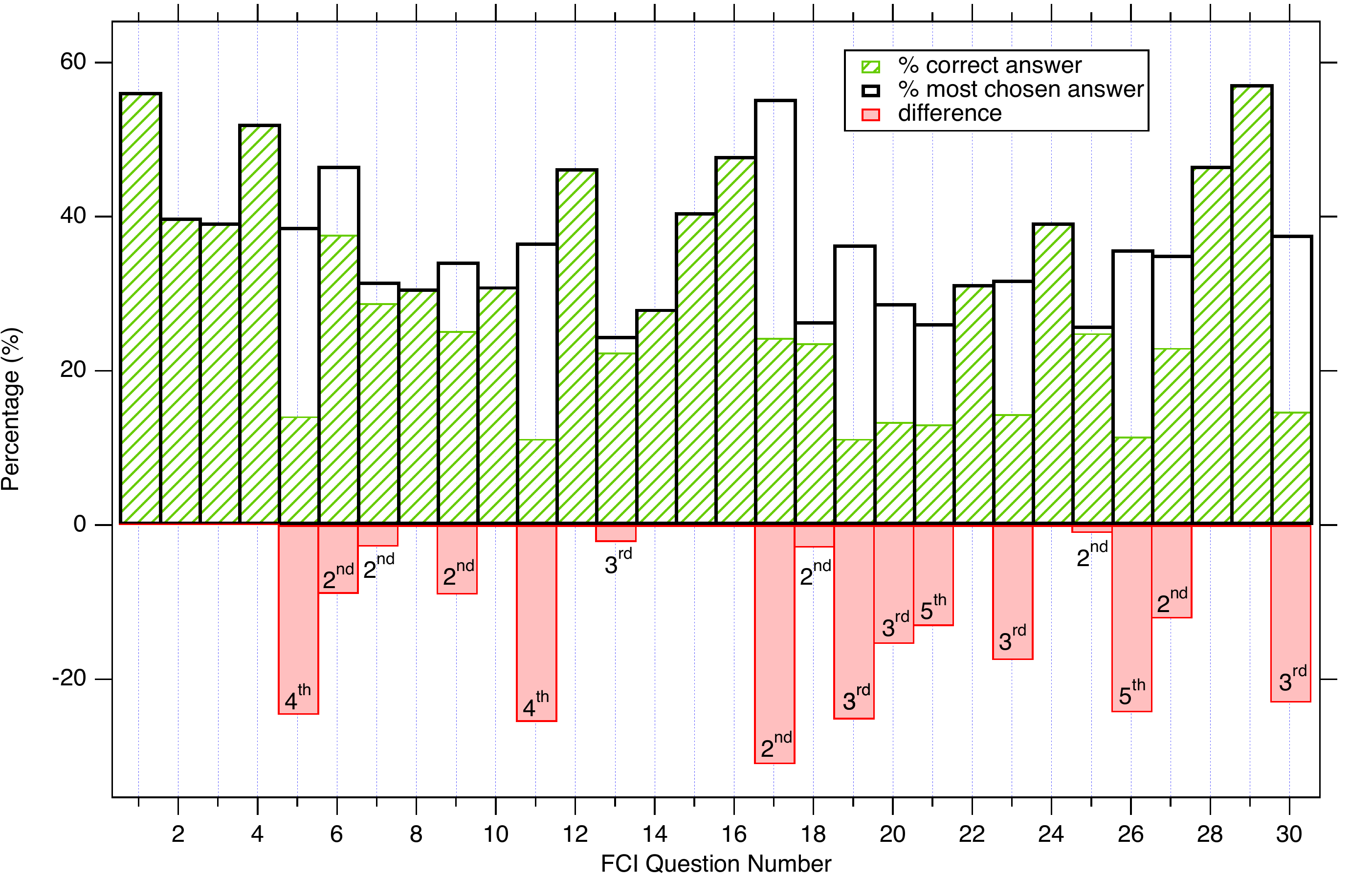}}
    \caption{Breakdown of pre-test questions for the 2021 University of Johannesburg students sitting an introductory physics course, $N=353$. The labels on the red \% differences, such as  $4^{th}, 2^{nd}$ etc., label the most commonly chosen (incorrect) response.}
    \label{fig:dom-misconcepts2021}
\end{figure}

  The preliminary analyses of the pre-tests for the 2021 cohort at the University of Johannesburg, for each of the 30 questions, is summarised in figure \ref{fig:dom-misconcepts2021}. The green shaded \% shows the percentage of students who got the correct answer, with the black outline showing the \% for the most commonly chosen answer. For example, questions 1-4 have the majority correct and this being the most chosen response. The choices where the majority response was incorrect correspond to those questions with a larger red \% difference showing that more students chose the wrong question $-$ for this cohort of students this corresponded to questions 5, 11, 17, 19, 26, and 30, respectively. 

\par These questions appear to show what Martin-Blas {\textit et al.} call ``dominant misconceptions"\footnote{Note that the majority wrong answer questions $-$  where the red shading \% is greater than the green shading \% $-$ lines up quite well with the ``polarising" questions \cite{Alinea:2015}: 5, 11, 13, 18, 29 and 30.} \cite{Martin_Blas:2010}. For example, question 5 focuses on circular motion and we can see here that this question was not well understood by most of the cohort. Given that this information was collected at the pre-test stage, the course instructors could attempt to adjust the teaching and learning to try to reinforce such a concept.

\par We will now turn our attention to the ``polarising" questions 5, 11, 13, 18, 29, and 30, which we observe to line up well with some of the dominant misconceptions found in figure \ref{fig:dom-misconcepts2021}. The responses to these questions are summarised in table \ref{table2} for each of the six classes. The interest in these particular questions lies in the fact that the ``polarising answer" is mostly correct but contains within it one intentionally misleading statement. For example, question 5 asks students to identify the forces driving an object's circular motion. The correct answer is B; the polarising choice of D lists the same forces as B but also includes ``a force in the direction of motion". As mentioned in section \ref{sec:1}, the polarising answer therefore reveals the students' misconception.  

\par We see this polarising effect to be most pronounced in classes where physics comprehension is assumed to be greatest $-$ the physics and life science majors, where the course prerequisites are higher, as well as the third-semester life sciences students, who have had the most exposure to Newtonian mechanics. However, it is concerning that the proportion of students polarised towards the wrong choice remains fairly high and even overtakes the correct answer (especially for questions 13 and 30, where more of these students favour the polarising rather than the correct answer).

\par The polarising effect remains fairly consistent for questions 18, 29, and 30 within the other studied classes; for questions 5, 11, and 13, the answers are far more scattered. These questions each concern motion in a frictionless environment, in the absence of a contact force. Again, it became apparent students struggled to understand the forces at play when objects are in motion. 


\section{Concluding remarks \label{sec:4}}

\par In these proceedings, we have used the pre-test part of the FCI to evaluate 
the baseline comprehension of Newtonian mechanics presented to first-year students 
enrolled in six introductory physics courses. We have demonstrated that 
misconceptions are more easily identified in students with a higher mechanics 
aptitude (considered to be the physics and life sciences majors in this study), but
remain persistent despite extensive training (as in the case of the life sciences 
students in the third semester of the extended programme). 

\par Although this is only a preliminary study, we hope to extend our data taking across future years (to eliminate any transient biases which may exist in this current cohort), where it would also be interesting to look at data not just from the University of Johannesburg, but also to extend the research by including other universities in South Africa that have similar demographics.

\section*{Acknowledgements}
 EC and ASC are supported in part by the National Research Foundation of 
South Africa (NRF). AC is  supported by the NRF and Department of Science and Innovation through the  SA-CERN programme. WN would like to thank Margaret Marshman (University of the Sunshine Coast, Australia) for useful comments and suggestions. 





\section*{References}


\end{document}